\documentclass[sigconf,nonacm]{acmart}
\pdfoutput=1
\setcopyright{cc}
\setcctype[4.0]{by}
\AtBeginDocument{%
  \providecommand\BibTeX{{%
    \normalfont B\kern-0.5em{\scshape i\kern-0.25em b}\kern-0.8em\TeX}}}
\acmConference[Scalesys '25]{1st International Workshop on Intelligent and Scalable Systems across the Computing Continuum }{November, 2025}{Vienna, Austria}
\usepackage[T1]{fontenc}
\usepackage{graphicx}
\usepackage{placeins}
\usepackage{float}
\usepackage{nomencl}
\usepackage{balance} 
\usepackage{color,soul}
\usepackage{multirow}
\usepackage{makecell}
\usepackage{longtable}
\usepackage{hhline}
\usepackage{pifont}
\usepackage{tabularx}
\usepackage[inline,shortlabels]{enumitem}
\usepackage{xspace}
\usepackage[binary-units,per-mode=symbol,exponent-product=\cdot,list-units=single,list-final-separator={,}]{siunitx}
\DeclareSIUnit[number-unit-product = ]\pixel{p}
\usepackage{chemformula}
\usepackage{gensymb}
\usepackage{amsmath}
\usepackage{booktabs}
\usepackage{comment}
\usepackage{footnote}
\usepackage{tablefootnote}
\usepackage{caption}
\usepackage{subcaption} 
\usepackage{wrapfig}
\usepackage[linesnumbered,ruled,vlined]{algorithm2e}
\usepackage[noend]{algpseudocode}
\usepackage{booktabs}
\usepackage{pgfplots}
\pgfplotsset{compat=1.18}
\usepackage{tikz}
\usepackage{amsmath}
\usepackage{xurl}
\usepackage{orcidlink}
\usepackage{hyperref}
\usepackage{hyperxmp}
\usepackage{footmisc}
\hypersetup{pdfauthor=author}
\begin{document}
\title{Toward Sustainability-Aware  LLM Inference on Edge Clusters}
\author{\mbox{Kolichala Rajashekar, Nafiseh Sharghivand, Radu Prodan}}
\affiliation{
  \institution{Department of Computer Science\\ University of Innsbruck, Austria}
  \city{}
  \country{}}
\author{Reza Farahani}
\affiliation{
  \institution{Institute of Information Technology\\ University of Klagenfurt, Austria}
  \city{}
  \country{}}

\begin{abstract}
Large language models (LLMs) require substantial computational resources, leading to significant carbon emissions and operational costs. Although training is energy-intensive, the long-term environmental burden arises from inference, amplified by the massive global query volume. Cloud-based inference offers scalability but suffers from latency and bandwidth constraints due to centralized processing and continuous data transfer. Edge clusters instead can mitigate these limitations by enabling localized execution, yet they face trade-offs between performance, energy efficiency, and device constraints. This short paper presents a sustainability-aware LLM inference for edge clusters comprising NVIDIA Jetson Orin NX (8GB) and Nvidia Ada 2000 (16GB) devices. It aims to balance inference latency and carbon footprint through carbon- and latency-aware routing strategies, guided by empirical benchmarking of energy consumption and execution time across diverse prompts and batch (i.e., group of prompts) configurations.
We compared baseline greedy strategies to carbon-aware and latency-aware strategies in prompt routing to specific hardware based on benchmarking information. 
Experimental evaluation shows that a batch size of four prompts achieves a trade-off between throughput, energy efficiency, while larger batches risk GPU memory saturation. 
\end{abstract}
\maketitle
\section{Introduction}
\label{sec:intro}
Large language models (LLMs) require immense computational resources, driving both high carbon emissions and substantial operational costs. Although LLM training is notoriously energy-intensive, the inference phase, where models process user prompts into responses, poses a greater long-term sustainability challenge due to the massive, continuous global query volume~\cite{azimi2025towards, varangot2025doing, fu2024llmco2}. Cloud servers provide vast computational and memory resources; however, processing all inference in the cloud requires extensive data transmission, which can degrade real-time performance under varying bandwidth availability~\cite {farahaniserverless, yang2024perllm}. Moreover, inference for large-scale models such as GPT-4 can emit, on a daily basis, a carbon footprint comparable to a significant fraction of their one-time training emissions, intensifying sustainability concerns amid accelerating AI adoption~\cite{fu2024llmco2,farahani2024towards}. To address these challenges, modern computing architectures increasingly integrate centralized cloud resources with distributed edge devices, where edge instances handle latency-sensitive and lightweight tasks locally, reducing response time and associated emissions, while cloud systems manage compute-intensive workloads at scale~\cite{farahani2025energyless, farahani2024heftless, li2024eaco}.

This hybrid paradigm is particularly crucial for LLM inference, where prompt complexity varies widely. For example, simple factual questions may require only modest computational effort, while tasks involving multi-step reasoning or tool use demand substantially greater processing power. However, current LLM inference systems still rely on coarse-grained heuristics, e.g., routing all reasoning-heavy or mathematical prompts to high-capacity models, ignoring hardware heterogeneity, inter-device communication latency, or dynamic energy profiles~\cite{kassem2025robust, chen2024routerdc, ding2024hybrid}. 
These oversights lead to inefficient resource utilization, elevated carbon emissions, and degraded performance in edge deployments.

This short paper introduces sustainable LLM inference on edge clusters composed of NVIDIA Jetson Orin NX (8GB) and Nvidia Ada 2000 (16GB) as representative edge hardware. We propose carbon- and latency-aware strategies informed by extensive benchmarking across diverse LLM prompts from established prompt datasets~\footnote{https://huggingface.co/datasets}. Through empirical evaluation of energy consumption, carbon emissions, and end-to-end latency, our results show that these strategies reduce emissions by up to \qty{35}{\percent} and improve execution speed by 2–3x compared to greedy baselines. Furthermore, we analyze the effect of batch size, i.e., the number of prompts processed in parallel during a single inference pass to amortize computational overhead, across configurations of \num{1}, \num{4}, and \num{8}. Experimental results shows that batch size of \num{4} provides the balance between end-to-end latency and energy efficiency, while a batch size of \num{8} increases GPU utilization at the cost of higher latency. These findings highlight the importance of benchmarking-driven, sustainability-aware LLM inference to enable efficient and environmentally responsible deployments on edge server clusters.
\section{Motivation Example}
\label{sec:motiv}
To illustrate the need for sustainability-aware LLM inference, we deployed two quantized Gemma models on an edge cluster: Gemma-3-1B-it-qat on an NVIDIA Jetson Orin NX (8GB) of GPU memory and Gemma-3-12B-it-qat on an NVIDIA Ada 2000 (16GB). For a cloud baseline, we used the Google Gemini 2.0 Flash API~\cite{team2025gemma}. We used Ollama and evaluated four representative prompts (P1–P4), summarized in Table~\ref{tab:prompts}, which cover reasoning, generative writing, and factual lookup. We used a judge model (cloud) that rates expected reasoning depth and token footprint, calculating prompt complexity scores (CS); scores are normalized to [0,1], higher is harder.
\begin{table}[t]
  \centering
  \footnotesize
  \caption{Prompts used in evaluation (CS: complexity score [0–1] from a judge model; higher is harder).}
  \label{tab:prompts}
  \begin{tabular}{|c|p{6.5cm}|c|}
    \hline
    \textbf{ID} & \textbf{Prompt} & \textbf{CS} \\
    \hline
    P1 & A group of five friends (Alice, Bob, Carol, David, Emily) are trying to decide who will buy tickets for a concert, prepare snacks, drive, and pick up drinks. Alice hates driving. Bob can only pick up drinks if he's not preparing snacks. Carol loves concerts and wants to buy tickets. David can only drive if Emily prepares snacks. Emily will not pick up drinks. Each friend must take exactly one task, and each task must be assigned to exactly one friend. Assign the tasks to each friend and explain your logical deduction step by step. & 0.47 \\
    \hline
    P2 & Write a short story, approximately 500 words, about a sentient, self-repairing antique grandfather clock that secretly orchestrates minor, benevolent 'time anomalies' in a quiet, forgotten library. Introduce a skeptical new librarian who slowly uncovers the clock's secret. The story must include: The clock's motivation for its actions. Three distinct 'time anomalies' are caused. A moment of direct, non-verbal communication between the clock and the librarian. A surprising twist where the librarian, instead of exposing the clock, aids its efforts for an unexpected reason. & 0.39 \\
    \hline
    P3 & What is the boiling point of water at standard atmospheric pressure? & 0.08 \\
    \hline
    P4 & Who painted the Mona Lisa? & 0.07 \\
    \hline
  \end{tabular}
\end{table}

\begin{figure}
\centering
\includegraphics[scale=0.49]{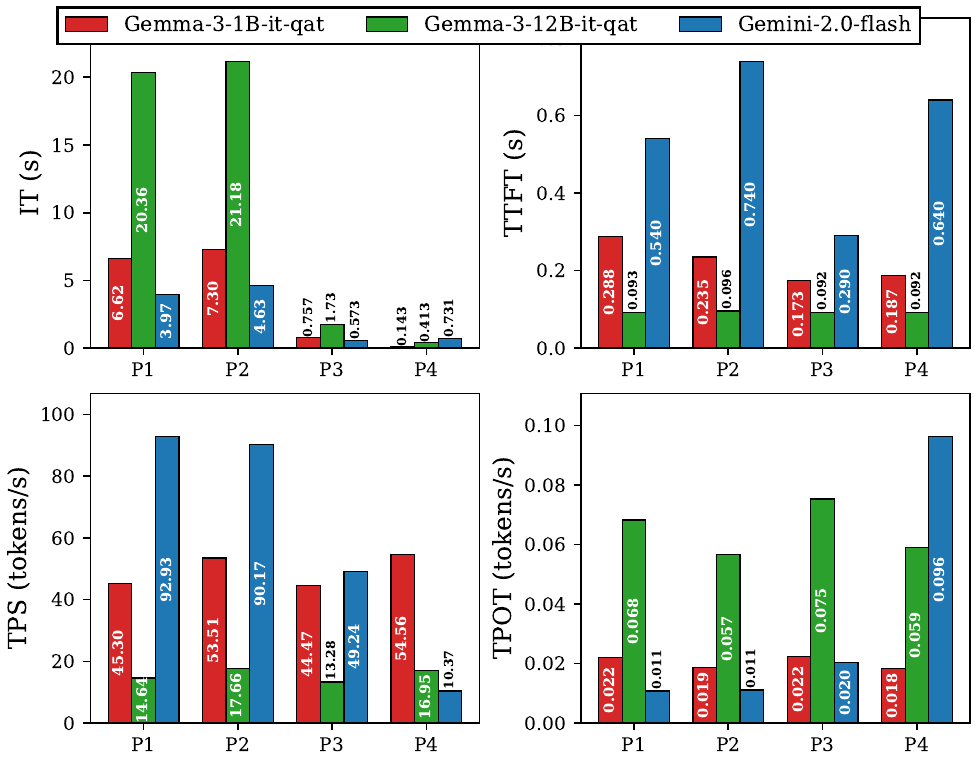}
\caption{Comparison of inference performance metrics, IT, TTFT, TPS, and TPOT, across NVIDIA Jetson Orin NX (8 GB), Ada 2000 (16 GB), and the Gemini 2.0 Flash cloud API.}
\label{fig:inference_time}
\end{figure}

\textit{Performance analysis}: We compare the performance of deployed LLM models using four key metrics: \textit{inference time} (IT), \textit{time-to-first-token} (TTFT), \textit{tokens-per-second} (TPS), and \textit{time-per-output-token} (TPOT)~\cite{behdin2025efficient, jain2024intelligent}. 
As shown in Fig.~\ref{fig:inference_time}, the Gemma-3-12B model achieves the shortest TTFT but incurs higher IT and TPOT on longer prompts, whereas Gemma-3-1B provides a more balanced efficiency profile. 
The cloud-based Gemini 2.0 Flash API delivers superior IT and TPS for complex prompts (P1, P2) but underperforms on simpler factual queries (P4), indicating bandwidth and dispatch overheads.

\begin{figure}[!t]
\footnotesize
    \centering
    \begin{subfigure}{\linewidth}
        \centering
        \includegraphics[scale=0.53]{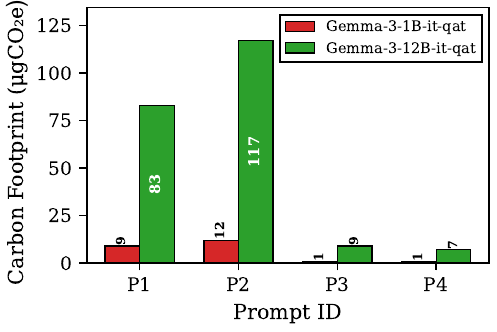}
        \label{fig:carbon}
    \end{subfigure}
    \vspace{-0.5em} 
    \begin{subfigure}{\linewidth}
        \centering
        \includegraphics[scale=0.53]{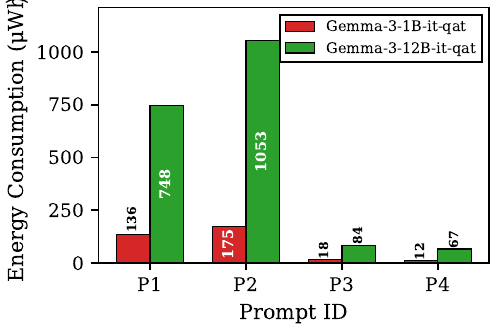}
        \label{fig:energy}
    \end{subfigure}
    \caption{Carbon footprint and energy consumption for Gemma3-1B-it and Gemma3-12B-it models across prompts P1–P4.}
    \label{fig:carbon-energy-combined}
\end{figure}

\textit{Sustainability analysis}: Fig.~\ref{fig:carbon-energy-combined} shows the measured carbon footprint (in $\mathrm{CO_2eq}$) and power draw (in watts) obtained using the JetPack SDK~\footnote{\url{https://developer.nvidia.com/embedded/jetpack}} and the PyNVML library~\footnote{\url{https://pypi.org/project/pynvml/}}. 
Gemma-3-1B emits roughly one-tenth the carbon of Gemma-3-12B on reasoning prompts (P1, P2), while both models exhibit low emissions on simpler ones (P3, P4). 
These results demonstrate that hardware-aware and model-adaptive inference can substantially reduce energy consumption and carbon emissions, underscoring the potential of sustainability-oriented deployment strategies for LLMs on edge clusters.


\textit{Key takeaway}: Sustainability-aware LLM inference on edge clusters demands balancing performance, energy, and carbon efficiency. Relying solely on either compact edge models or large cloud-based LLMs is suboptimal.  Instead, hardware- and model-aware LLM inference is crucial for minimizing emissions while maintaining responsiveness. As shown in Fig.~\ref{fig:inference_time}, lightweight models such as Gemma-3-1B achieve low latency for simple prompts, whereas Fig.~\ref{fig:carbon-energy-combined} confirms substantial energy and carbon savings. For complex reasoning tasks, larger models, such as Gemma-3-12B or the Gemini 2.0 Flash API, offer superior output quality, underscoring the need for workload distribution across the edge–cloud continuum.
\section{Benchmark Evaluation}
We assessed our edge cluster, introduced in Section~\ref{sec:motiv}, using a mixed dataset that integrates prompts from multiple publicly available sources. 
The prompts spans diverse domains, including math reasoning (GSM8K)~\cite{cobbe2021gsm8k}, extractive question answering (SQuAD)~\cite{rajpurkar-etal-2016-squad}, dialogue summarization (DialogSum)~\cite{chen-etal-2021-dialogsum}, Python coding instructions~\cite{bisht2021pythoncodeinstructions}, multiple-choice science reasoning (ARC-Challenge)~\cite{allenai:arc}, long-form summarization of arXiv papers, multi-turn dialogue continuation~\cite{li2017dailydialog}, and general long-form summarization~\cite{hermann2015cnndm}. From this composite benchmark of approximately \num{5000} prompts, we sampled \num{500} representative inputs to measure both end-to-end inference latency and the corresponding carbon footprint across the edge cluster. Table \ref{tab:performance_metrics} summarizes the benchmark results, reporting average performance metrics across all evaluated configurations. These results provide an overview of the latency–energy trade-offs and serve as a practical guide for resource allocation in edge–cloud deployments. We implemented two LLM inference strategies:
\begin{table*}[h]
\footnotesize
\caption{Average inference metrics across edge devices and batch configurations.}
\footnotesize
\label{tab:performance_metrics}
\centering
\begin{tabular}{|l|c|c|c|c|c|c|c|c|}
\hline
\textbf{Hardware} & \textbf{Batch Size} & \textbf{E2E Latency (s)} & \textbf{ TTFT (s)} & \textbf{TPOT (s)} & \textbf{Token Count} & \textbf{Tokens/s (Throughput)} & \textbf{Energy (kWh)} & \textbf{Carbon (kgCO2e)} \\
\hline
& 1 & 3.39 & 0.26 & 0.03 & 69.62 & 20.54 & 6.35e-05 & 4.38e-06 \\
Ada 2000 16GB & 4 & 14.58 & 12.07 & 0.02 & 56.83 & 3.90 & 5.05e-05 & 3.49e-06 \\
& 8 & 26.82 & 24.00 & 0.03 & 63.97 & 2.39 & 5.73e-05 & 3.96e-06 \\
\hline
& 1 & 13.06 & 0.36 & 0.061 & 148 & 11.33 & 1.79e-05 & 1.23e-06 \\
Jetson 8GB & 4 & 15.08 & 1.13 & 0.063 & 149 & 9.88 & 4.89e-06 & 3.37e-07 \\
& 8 & 14.12 & 4.87 & 0.057 & 136 & 9.63 & 5.12e-06 & 3.53e-07 \\
\hline
\end{tabular}
\end{table*}

\textit{(i) Carbon-aware}: Assigns each prompt to the model with the measured lower carbon footprint, prioritizing emission reduction even if it increases latency.

\textit{(ii) Latency-aware}: Employs a greedy heuristic that sorts prompts by decreasing average latency and assigns them to minimize total end-to-end execution time.

Baselines include assigning all prompts exclusively to either the Jetson Orin NX (8 GB) or the Ada 2000 (16 GB). The LLM inference results for batch sizes of \num{1}, \num{4}, and \num{8}, summarized in Table~\ref{tab:results}, reveal clear trade-offs between execution time and carbon footprint. For a batch size of \num{1}, assigning all prompts to the NVIDIA Jetson (8GB) yields a total execution time of \qty{1873.13}{\second} with a carbon footprint of 0.000209 kg CO\textsubscript{2}e, whereas using the NVIDIA Ada (16GB) reduces execution time to \qty{1354.25}{\second} but increases emissions to 0.000300 kg CO\textsubscript{2}e. The \textit{carbon-aware} strategy achieves the minimum footprint by directing roughly \qty{85}{\percent} of prompts to the Jetson device, 
leveraging its energy efficiency for low-token tasks such as sentiment analysis. However, this causes high end-to-end (E2E) latency due to load imbalance from compute-intensive tasks such as Python coding.  
In contrast, the \textit{latency-aware} strategy minimizes total E2E latency to \qty{580.34}{\second} by balancing workload distribution, assigning complex tasks to the Ada device, while maintaining a moderate carbon footprint of 0.000247 kg CO\textsubscript{2}e. These results highlight the complementary roles of both devices: the Jotson device excels in lightweight workloads, whereas the Ada instance dominates in memory- and compute-intensive inference.

For a batch size of \num{4}, the \textit{Jetson-only} baseline achieves an execution time of \qty{649.6}{\second} with a carbon footprint of 0.000071 kg CO\textsubscript{2}e, while the \textit{Ada-only} configuration reduces execution time to \qty{568.4}{\second} but increases emissions to 0.000103 kg CO\textsubscript{2}e. The \textit{carbon-aware} strategy lowers emissions by approximately \qty{33}{\percent} (0.000069 kg CO\textsubscript{2}e) by routing around \qty{80}{\percent} of prompts to the Ada device. However, E2E latency remains high due to memory constraints affecting compute-intensive tasks such as Python coding. In contrast, the \textit{latency-aware} strategy achieves the shortest E2E latency, about twice as fast as the \textit{Jetson-only} baseline, while maintaining a moderate carbon footprint of 0.000085 kg CO\textsubscript{2}e, benefiting from balanced workload assignment that exploits the Ada device’s efficiency for high-token prompts. Overall, a batch size of \num{4} provides a strong trade-off between throughput and energy efficiency, although minor accuracy degradation on the Ada device indicates limitations in handling larger model states.

At a batch size of \num{8}, the \textit{Jetson-only} baseline shows an execution time of \qty{609}{\second} with a carbon footprint of 0.000057 kg CO\textsubscript{2}e, while the \textit{Ada-only} setup reduces execution time to  \qty{533.6}{\second} but increases emissions to 0.000084 kg CO\textsubscript{2}e. The \textit{carbon-aware} strategy yields the lowest footprint (0.000055 kg CO\textsubscript{2}e) by routing approximately \qty{75}{\percent} of prompts to the Jetson device. However, its E2E latency (\qty{552.4}{\second}) remains elevated due to instability on high-token workloads. In contrast, the \textit{latency-aware} strategy minimizes E2E latency to \qty{266.8}{\second}, roughly twice as fast as the \textit{Jetson-only} baseline, while maintaining a moderate footprint of 0.000070 kg CO\textsubscript{2}e, benefiting from the Ada device’s greater stability in long-form summarization and other memory-intensive tasks. Although a batch size of \num{8} maximizes throughput, it introduces instability and accuracy degradation on the Jetson device, indicating that larger-memory configurations are preferable for high-batch inference workloads.
\begin{table}[htbp]
    \caption{Comparison of different LLM inference strategies across batch sizes 1, 4, and 8.}
    \footnotesize
    \centering
    \begin{tabular}{lcc}
        \hline
        \textbf{Strategy} & \textbf{Total E2E latency (s)} & \textbf{Total Carbon Footprint (kgCO2e)} \\
        \hline
        \multicolumn{3}{c}{\textbf{Batch Size 1}} \\
        \hline
        All on Jetson (8GB) & 1873.13 & 0.000209 \\
        \hline
        All on Ada (16GB) & 1354.25 & 0.000300 \\
        \hline
        Carbon-Aware & 1674.86 & 0.000204 (lowest) \\
        \hline
        Latency-Aware & 580.34 (lowest) & 0.000247 \\
        \hline
        \multicolumn{3}{c}{\textbf{Batch Size 4}} \\
        \hline
        All on Jetson (8GB) & 649.6 & 0.000071 \\
        \hline
        All on Ada (16GB) & 568.4 & 0.000103 \\
        \hline
        Carbon-Aware & 590.2 & 0.000069 (lowest) \\
        \hline
        Latency-Aware & 284.2 (lowest) & 0.000085 \\
        \hline
        \multicolumn{3}{c}{\textbf{Batch Size 8}} \\
        \hline
        All on Jetson (8GB) & 609.0 & 0.000057 \\
        \hline
        All on Ada (16GB) & 533.6 & 0.000084 \\
        \hline
        Carbon-Aware & 552.4 & 0.000055 (lowest) \\
        \hline
        Latency-Aware & 266.8 (lowest) & 0.000070 \\
        \hline
    \end{tabular}
    \label{tab:results}
\end{table}

Cross-batch analysis reveals that overall latency decreases with larger batch size, as parallel token generation reduces
TPOT. However, TTFT increases significantly, noticeably, limiting responsiveness in real-time applications. The carbon footprint per prompt declines with batching, since energy costs are amortized across multiple inputs.  
The Ada device consistently achieves higher accuracy, particularly at batch size \num{8}, where the Jetson device exhibits errors due to memory saturation. Overall, batch size \num{4} offers the best trade-off between latency, carbon footprint, and accuracy, whereas batch size \num{8} maximizes throughput but demands at least 16GB of GPU memory for stable operation. The Jetson instance remains well-suited for lightweight, low-token tasks, while the Ada device excels in high-token, high-batch inference scenarios.
\section{Conclusion and Future Work}

This paper presented an empirical analysis of sustainability-aware LLM inference on heterogeneous edge server clusters. Our results show that increasing and balancing prompt batch sizes provides a clear trade-off between latency and carbon footprint, making it suitable for mixed workloads. In contrast, a larger batch size maximizes throughput but requires more memory for stability. In our experiments, the carbon-aware strategy reduces emissions by up to \qty{35}{\percent} by leveraging the Jetson device efficiency for low-token tasks, while the latency-aware strategy achieves 2-3x faster execution times through balanced load distribution.  These findings highlight the importance of dynamic, complexity-aware LLM inference for optimizing performance and minimizing environmental impact. Future work will investigate scalability for unseen prompts and adaptive edge-server selection to advance sustainable LLM inference.
\section*{Acknowledgment}
This work received funding from the Horizon Europe research and innovation program  (grant agreement 101189771, DataPACT) and the Austrian Research Promotion Agency (FFG, grant agreement 909989, AIM AT Stiftungsprofessur für Edge AI).
\bibliographystyle{ACM-Reference-Format}
\bibliography{main}

\end{document}